\documentclass{article}
\usepackage{spconf,amsmath,graphicx}
\usepackage{cite}
\usepackage{color}
\usepackage{tabularx}
\usepackage{mathrsfs,amsmath,amsfonts}
\usepackage{graphicx}
\usepackage[T1]{fontenc}
\usepackage{hyperref}
\hypersetup{hidelinks}
\usepackage{lipsum}
\usepackage{booktabs}

\usepackage{tikz}
\usepackage[mode=buildnew]{standalone}
\usetikzlibrary{dsp}
\usetikzlibrary{chains}
\usetikzlibrary{shapes.geometric}
\usetikzlibrary{calc}

\usepackage{pgfplots}
    \pgfplotsset{compat=1.8}

\pgfplotsset{
    /pgfplots/area legend/.style={
        /pgfplots/legend image code/.code={
            \fill[##1] (0cm,0.6em) rectangle (0.9em,-0.3em);
}, },
}

\usepackage[input-decimal-markers={,}, 
     output-decimal-marker = {,},]{siunitx}
\let\OLDthebibliography\thebibliography
\renewcommand\thebibliography[1]{
  \OLDthebibliography{#1}
  \setlength{\parskip}{0.5pt}
  \setlength{\itemsep}{0.32pt plus 5ex}
}

\newcommand{\erfc}{{\rm erfc}}

\newcommand{\putindex}[3]{\vtop{\hbox{\hspace{#3} $#1$}
            \hbox{\raise 6mm \hbox{$\scriptscriptstyle #2$}}}}

\newcommand{\gradx}[0]{\vtop{\hbox{\rm grad}
            \hbox{\raise 2.5mm \hbox{\rm \hspace{2mm} \footnotesize x}}}}

\newcommand{\grady}[0]{\vtop{\hbox{\rm grad}
            \hbox{\raise 2.5mm \hbox{\rm \hspace{2mm} \footnotesize y}}}}

\newcommand{\grad}[1]{\vtop{\hbox{\rm grad}
            \hbox{\raise 2.5mm \hbox{#1}}}}

\newcommand{\btb}{     \begin{tabbing}             }
\newcommand{\bte}{     \end{tabbing}               }

\title{EFFICIENT HIGH-PERFORMANCE BARK-SCALE NEURAL NETWORK \\ FOR RESIDUAL ECHO AND NOISE SUPPRESSION}

\name{Ernst Seidel$^{\ast}$, \it Pejman Mowlaee$^{\circ}$, Tim Fingscheidt$^{\ast}$}
\address{$^{\ast}$Institute for Communications Technology, Technische Universit\"{a}t Braunschweig\\	Schleinitzstra{\ss}e 22,	38106 Braunschweig, Germany\\ $^{\circ}$GN Audio A/S,	Lautrupbjerg 7,	2750 Ballerup, Denmark}
\begin{document}
\ninept
\maketitle
\begin{abstract}
In recent years, the introduction of neural networks (NNs) into the field of speech enhancement has brought significant improvements. However, many of the proposed methods are quite demanding in terms of computational complexity and memory footprint. For the application in dedicated communication devices, such as speakerphones, hands-free car systems, or smartphones, efficiency plays a major role along with performance. In this context, we present an efficient, high-performance hybrid joint acoustic echo control and noise suppression system, whereby our main contribution is the postfilter NN, performing both noise and residual echo suppression. The preservation of nearend speech is improved by a Bark-scale auditory filterbank for the NN postfilter. The proposed hybrid method is benchmarked with state-of-the-art methods and its effectiveness is demonstrated on the ICASSP 2023 AEC Challenge blind test set. We demonstrate that it offers high-quality nearend speech preservation during both double-talk and nearend speech conditions. At the same time, it is capable of efficient removal of echo leaks, achieving a comparable performance to already small state-of-the-art models such as the end-to-end DeepVQE-S, while requiring only around 10\% of its computational complexity. This makes it easily realtime implementable on a speakerphone device.
\end{abstract}
\vspace{-0.2cm}
\section{Introduction}
\label{sec:intro}

Echo and background noise are major obstacles in everyday life's speech communication. To enable high-quality speakerphones or video conference solutions, a reliable and efficient joint denoising and acoustic echo control solution is strictly required. In particular, as a remaining major challenge for speech communication devices used in real-life, double-talk performance has been often reported as a limiting factor, as the device shall allow users to speak and hear simultaneously without interruptions. Furthermore, a reduced double-talk performance was shown to impair meeting inclusiveness and participation rate \cite{Cutler2021MeetingEA} \footnote{We would like to thank Rasmus K. Olsson for his fruitful discussion}.\\ 
\indent Recently, there has been quite some interest towards deep learning-based solutions for acoustic echo control. Various methods relying on a neural network (NN) have also been proposed for the task of joint denoising and echo cancellation. They can be categorized into two groups: 
(1) \textit{Fully learned methods}, which describe either a two-stage NN approach with a dedicated acoustic echo control module and a second noise and residual echo suppression (RES) module \cite{seidel21_interspeech,Franzen2021,Braun_IWAENC2022,Seidel_IWAENC2022,Seidel_WASPAA2023}, or a single NN trained for a joint task, as an example dereverberation, denoising and echo cancellation (see, e.g., \cite{indenbom2023deepvqe,sca,indenbom2023deep,westhausen21_icassp,Wang2022}).
(2) {\textit{Hybrid methods}} are themselves grouped into two categories: \mbox{(i) a combined} linear echo canceller (LEC) \cite{Enzner2006,steinert_lowdelayhandsfree} followed by an NN postfilter (PF) as residual echo and/or noise suppressor \cite{panchapagesan23_interspeech,pfeifenberger21_interspeech,Shachar2023,Valin_AEC}, or  (ii) {an NN-aided step-size control or state estimation for the LEC \cite{Haubner2021,NKF2023}}.\\
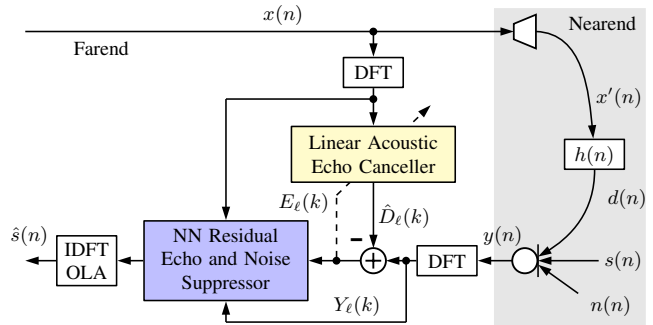
\begin{figure}[t]
\centering    
\hspace{-4.3mm}
	\usetikzlibrary{dsp}
	\usetikzlibrary{chains}
	\usetikzlibrary{shapes.geometric}
	\usetikzlibrary{calc}

\begin{tikzpicture}[scale = 1.3]

\tikzstyle{encdec}=[trapezium, trapezium angle=67.5, draw, inner ysep=5pt, outer sep=0pt,text width=1, minimum height=10, line width=.8pt, fill=white]

\pgfarrowsdeclare{dsparrow}{dsparrow}
{
	\arrowsize=0.40pt
	\advance\arrowsize by .5\pgflinewidth
	\pgfarrowsleftextend{-4\arrowsize}
	\pgfarrowsrightextend{4\arrowsize}
}
{
	\arrowsize=0.40pt
	\advance\arrowsize by .5\pgflinewidth
	\pgfsetdash{}{0pt} 
	\pgfsetmiterjoin	 
	\pgfsetbuttcap		 
	\pgfpathmoveto{\pgfpoint{-4\arrowsize}{2.5\arrowsize}}
	\pgfpathlineto{\pgfpoint{4\arrowsize}{0pt}}
	\pgfpathlineto{\pgfpoint{-4\arrowsize}{-2.5\arrowsize}}
	\pgfpathclose
	\pgfusepathqfill
}

\pgfmathsetmacro{\toprow}{3.8}
\pgfmathsetmacro{\bottomrow}{1}
\pgfmathsetmacro{\middlerow}{(2.30}
\pgfmathsetmacro{\boxheight}{2}
\pgfmathsetmacro{\boxwidth}{5.5}

    \draw (4.05, \bottomrow) node  (e_bot) [dspnodefull] {}
            (4.05, \middlerow-0.5) coordinate  (e_mid)  {}
            (5.2, \middlerow+0.6) coordinate  (e_top)  {};

    \begin{scope}[start chain, dashed]

        \chainin (e_bot);
        \chainin (e_mid)  [join=by dspline];
        \chainin (e_top)  [join=by dspconn];
        
    \end{scope}

    \draw   (6.9, \middlerow-0.15) node    (backdrop)     [dspfilter, minimum width=7.7 em, minimum height=16.0 em, fill=black!10, draw=white] {}
            (7.3, 3.9) node (Sys) {Nearend}
            (1.18, 3.6) node (Sys) {Farend}
            (0.25, \toprow)    coordinate  (InX)   {}
            (4.5, \toprow)    node  (StopX2) [dspnodefull] {}

            (4.5, \toprow-0.5)    node  (AFB2) [dspfilter,  minimum width=3 em, minimum height=1.5 em] {DFT}
            (5.4, \bottomrow)    node  (AFB3) [dspfilter,  minimum width=3 em, minimum height=1.5 em] {DFT}
            (1.0, \bottomrow)    node  (SFB) [dspfilter,  minimum width=3 em, minimum height=2.5 em] {}
            (1.0, \bottomrow)    node  (text) [align = center, text width = 3em] {IDFT OLA}

            (4.5, \toprow-0.825)    node  (split) [dspnodefull] {}
            (2.7, \toprow-0.825)    coordinate  (split2) {}
            


            (6.35, \toprow)  node (LS) [encdec, rotate=90] {}

            (7.2, \middlerow)     node (MidD)   [dspfilter, minimum width=3 em, minimum height=1.5 em, fill=white] {$h(n)$}
            (7.25, \bottomrow)     coordinate (S)      {}
            (7.0, \bottomrow-0.4)   coordinate (N)      {}

            (4.5, \middlerow+0.0) node        (LEC) [dspfilter, minimum width=2.6cm, minimum height=1.4*\boxheight em, fill=yellow!25] {}
            (4.5, \middlerow+0.0) node (text)  [text width=3cm, align=center]{Linear Acoustic Echo Canceller} 
            (4.5, \bottomrow) node  (add) [dspadder, minimum width=0.5, fill=white]  {}

            (4.9, \bottomrow) node  (bp1) [dspnodefull] {}
            (4.9, \bottomrow-0.75) coordinate  (bp2)  {}
            (2.7, \bottomrow-0.75) coordinate  (bp3)  {}
            
            (0.25, \bottomrow) coordinate  (OutE)  {}
            (2.7, \bottomrow) node        (PF) [dspfilter, minimum width=2.6cm, minimum height=2*\boxheight em, fill=blue!25] {}
            (2.7, \bottomrow) node (text)  [text width=2.5cm, align=center]{NN Residual Echo and Noise Suppressor} 
            (6.35, \bottomrow) node  (MicIn) [circle, inner sep= 4, fill=white]  {}
            (6.35, \bottomrow) node  (Mic) [dspmultiplier, minimum width=0.5, fill=white]  {}
            (6.51, \bottomrow) coordinate  (InM)  {}
            (6.51, \bottomrow+0.2) coordinate  (MT)  {}
            (6.51, \bottomrow-0.2) coordinate  (MB)  {};

    \draw (3.4, \toprow+0.2) node (Xt) {$x(n)$};
    \draw (0.3, \bottomrow+0.3) node (St) {$\hat{s}(n)$};
    \draw (3.65, \bottomrow+0.70) node (Et) {$E_\ell(k)$};
    \draw (4.875, \bottomrow+0.55) node (Dt) {$\hat{D}_\ell(k)$};
    \draw (4.3, \bottomrow+0.2) node (mint) {\huge -};
    \draw (4.3, \bottomrow-0.55) node (Yt) {$Y_\ell(k)$};
    \draw (6.08, \bottomrow+0.30) node (Yt) {$y(n)$};
    \draw ([xshift=+3mm, yshift=0mm] S.east)                node (St) {$s(n)$};
    \draw ([xshift=+4mm, yshift=-1.5mm] N.east)           node (Nt) {$n(n)$};
    \draw ([xshift=+4mm, yshift=-3.5mm] MidD.south)           node (Dt) {$d(n)$};
    \draw ([xshift=+3mm, yshift=+5mm] MidD.north)           node (X_t) {$x'(n)$};
    \draw ([yshift=-.5mm] InM) coordinate (InMb) {};

    \begin{scope}[start chain]
	
        \chainin (InX);
        \chainin (LS) 	[join=by dspconn];

        \chainin (Mic);
        \chainin (AFB3) 	[join=by dspconn];
        \chainin (add)      [join=by dspconn];
        \chainin (PF) 	[join=by dspconn];
        \chainin (SFB) 	[join=by dspconn];
        \chainin (OutE) 	[join=by dspconn];

        \chainin (StopX2);
        \chainin (AFB2) 	[join=by dspconn];
        \chainin (LEC) 	[join=by dspconn];
        \chainin (add) 	[join=by dspconn];

        \chainin (split);
        \chainin (split2) 	[join=by dspline];
        \chainin (PF) 	[join=by dspconn];

        \chainin (S);
        \chainin (InM) 	[join=by dspconn];
        \chainin (N);
        \chainin (InMb) 	[join=by dspconn];

        \chainin (bp1);
        \chainin (bp2) [join=by dspline];
        \chainin (bp3) [join=by dspline];
        \chainin (PF) [join=by dspconn];

        \chainin (MT);
        \chainin (MB) [join=by dspline];

    \end{scope}

    \draw[dspconn] (LS.south) to[out=-2,in=95] (MidD);
    \draw[dspconn] (MidD) to[out=-90,in=30] ([yshift=.5mm] InM);
	
\end{tikzpicture}
\vspace{-0.7cm}
\caption{Hybrid system consisting of a linear acoustic echo canceller and the \textbf{proposed neural postfilter performing residual echo and noise suppression}, details are shown in Fig.\,\ref{fig:Figure2}.}\label{fig:Figure1}
\vspace{-0.6cm}
\end{figure}
\indent Major contributions to the advancement of NN-based acoustic echo control have been the Acoustic Echo Cancellation Challenges organized by Microsoft \cite{cutler2023AEC}. Top-performing methods have been reported at these challenges, but their footprints and performance vs. complexity trade-offs are often not allowing for implementation in consumer devices, e.g., a speakerphone or a car's handsfree system. \\
\indent The recent challenges furthermore revealed that there is still headroom for improving acoustic echo control performance w.r.t.\ the metrics "double-talk nearend speech preservation" (DT Other) and "single-talk nearend (STNE) MOS" (ST Other) \cite{cutler2023AEC}. A possible solution could be the application of an auditory filterbank to imitate the frequency resolution of the human hearing system \cite{valin2020,Valin_AEC}. \\
\indent Motivated by the above facts and to address the existing challenges, we propose an efficient high-performance NN design for noise and residual echo suppression. The proposed method relies on a classical LEC, which is combined with a low-complexity NN PF incorporating a Bark-scale auditory filterbank. Through various experiments, we demonstrate that the proposed solution offers a good trade-off between echo/noise suppression and near-end speech preservation, while being realtime-capable in a speakerphone device due to its low complexity and memory footprint.\\
\indent The rest of this paper is organized as follows: In the following Section~\ref{sec:Method}, we present the proposed method and its details. Section~\ref{sec:Framework} elaborates on dataset, training framework, reference methods, and metrics employed in this work. In Section~\ref{sec:Results}, we present the results and discussions. Finally, Section~\ref{sec:Conclusion} concludes our work.

\section{Proposed Method}
\label{sec:Method}

\begin{figure}[t]
\centering    
\hspace{-4.0mm}
	\usetikzlibrary{dsp}
	\usetikzlibrary{chains}
	\usetikzlibrary{shapes.geometric}
	\usetikzlibrary{calc}

\begin{tikzpicture}[scale = 1.0]

\tikzstyle{encdec}=[trapezium, trapezium angle=67.5, draw, inner ysep=5pt, outer sep=0pt,text width=1, minimum height=10, line width=.8pt, fill=white]

\pgfarrowsdeclare{dsparrow}{dsparrow}
{
	\arrowsize=0.40pt
	\advance\arrowsize by .5\pgflinewidth
	\pgfarrowsleftextend{-4\arrowsize}
	\pgfarrowsrightextend{4\arrowsize}
}
{
	\arrowsize=0.40pt
	\advance\arrowsize by .5\pgflinewidth
	\pgfsetdash{}{0pt} 
	\pgfsetmiterjoin	 
	\pgfsetbuttcap		 
	\pgfpathmoveto{\pgfpoint{-4\arrowsize}{2.5\arrowsize}}
	\pgfpathlineto{\pgfpoint{4\arrowsize}{0pt}}
	\pgfpathlineto{\pgfpoint{-4\arrowsize}{-2.5\arrowsize}}
	\pgfpathclose
	\pgfusepathqfill
}

\pgfmathsetmacro{\toprow}{4.1}
\pgfmathsetmacro{\bottomrow}{1}
\pgfmathsetmacro{\middlerow}{(2.30}
\pgfmathsetmacro{\boxheight}{2}
\pgfmathsetmacro{\boxwidth}{5.5}

\pgfmathsetmacro{\vsc}{0.965}

    \draw (6.8*\vsc, \middlerow+0.25) node (back) [rectangle, minimum width=35.25 em *\vsc, minimum height=11.5 em, fill=blue!25] {};

    \draw (1.2*\vsc, \toprow) coordinate (ec0) {};
    \draw (1.2*\vsc, \middlerow) coordinate (y0) {};
    \draw (1.2*\vsc, \middlerow-1) coordinate (x0) {};

    \draw (2.60*\vsc, \middlerow+1) node (ecA) [dspfilter, minimum width=0.85cm, fill=white] {$|\cdot|^2$} ;
    \draw (2.60*\vsc, \middlerow) node (yA) [dspfilter, minimum width=0.85cm, fill=white] {$|\cdot|^2$} ;
    \draw (2.60*\vsc, \middlerow-1) node (xA) [dspfilter, minimum width=0.85cm, fill=white] {$|\cdot|^2$} ;

    \draw (3.775*\vsc, \middlerow+1) node (ecB) [dspfilter, dashed, minimum width=0.8cm, fill=green!] {$\boldsymbol{B}$}
    ++ (0.75 *\vsc, 0) coordinate (ecIn) {};
    \draw (3.775*\vsc, \middlerow) node (yB) [dspfilter, dashed, minimum width=0.8cm, fill=green!] {$\boldsymbol{B}$} ;
    \draw (3.775*\vsc, \middlerow-1) node (xB) [dspfilter, dashed, minimum width=0.8cm, fill=green!] {$\boldsymbol{B}$} 
    ++ (0.75 *\vsc, 0) coordinate (xIn) {};

    \draw (4.8*\vsc, \middlerow) node (conc) [dspfilter, fill=white, minimum width=0.55cm, minimum height=2.8cm] {} ;
    \draw (4.8*\vsc, \middlerow) node (text) {\rotatebox{90}{Concat / $\log(\cdot)$}} ;

    \draw (5.7*\vsc, \middlerow) node (nn1) [dspfilter, fill=white, minimum width=0.55cm, minimum height=2cm] {} ;
    \draw (5.7*\vsc, \middlerow) node (text) {\rotatebox{90}{FC($400$)}} ;

    \draw (6.6*\vsc, \middlerow) node (nn2) [dspfilter, fill=white, minimum width=0.55cm, minimum height=2cm] {} ;
    \draw (6.6*\vsc, \middlerow) node (text) {\rotatebox{90}{GRU($257$)}} ;

    \draw (7.5*\vsc, \middlerow) node (nn3) [dspfilter, fill=white, minimum width=0.556cm, minimum height=2cm] {} ;
    \draw (7.5*\vsc, \middlerow) node (text) {\rotatebox{90}{GRU($257$)}} ;

    \draw (8.4*\vsc, \middlerow) node (nn4) [dspfilter, fill=white, minimum width=0.55cm, minimum height=2cm] {} ;
    \draw (8.4*\vsc, \middlerow) node (text) {\rotatebox{90}{FC($600$)}} ;

    \draw (9.3*\vsc, \middlerow) node (nn5) [dspfilter, fill=white, minimum width=0.55cm, minimum height=2cm] {} ;
    \draw (9.3*\vsc, \middlerow) node (text) {\rotatebox{90}{FC($600$)}} ;

    \draw (10.2*\vsc, \middlerow) node (nn6) [dspfilter, fill=white, minimum width=0.55cm, minimum height=2cm] {} ;
    \draw (10.2*\vsc, \middlerow) node (text) {\rotatebox{90}{FC($86$)}} ;

    \draw (11.25*\vsc, \middlerow) node (mB) [dspfilter, dashed, fill=white, minimum width=0.8cm, fill=green!] {$\boldsymbol{B}^T$} ;

    \draw (11.85*\vsc, \middlerow) coordinate (oC) {} ;
    \draw (11.85*\vsc, \toprow) node (circ) [circle, fill=white, minimum size=1.1 em] {} ;
    \draw (11.85*\vsc, \toprow) node (mult) [dspmixer] {} ;

    \draw (1.5*\vsc, \middlerow+1) coordinate (ecS) {};
    \draw (1.5*\vsc, \toprow) node (bp1)  [dspnodefull] {} ;


    \draw (12.75*\vsc, \toprow) coordinate (shat) {} ;

    \draw ([xshift = 4mm, yshift = -11mm] ec0) node (ECt) {$E_\ell(k)$};
    \draw ([xshift = 4mm, yshift = -3mm] y0) node (Yt) {$Y_\ell(k)$};
    \draw ([xshift = 4mm, yshift = -3mm] x0) node (Xt) {$X_\ell(k)$};

    \draw ([xshift = -5mm, yshift = 10mm] oC) node (Mt) {$M_\ell(k)$};
    \draw ([xshift = -10mm, yshift = -2.5mm] mult) node (mt) {$E_\ell(k)$};
    \draw ([xshift = -3.5mm, yshift = -3mm] shat) node (St) {$\hat{S}_\ell(k)$};

    \draw (8.65*\vsc, 0.9) node (St) {NN Residual Echo and Noise Suppressor};

    \begin{scope}[start chain]

        \chainin (bp1);
        \chainin (ecS) [join= by dspline];
        \chainin (ecA) [join= by dspconn];
        \chainin (ecB) [join= by dspconn];
        \chainin (ecIn) [join= by dspconn];

        \chainin (x0);
        \chainin (xA) [join= by dspconn];
        \chainin (xB) [join= by dspconn];
        \chainin (xIn) [join= by dspconn];

        \chainin (y0);
        \chainin (yA) [join= by dspconn];
        \chainin (yB) [join= by dspconn];
        \chainin (conc) [join= by dspconn];
        \chainin (nn1) [join= by dspconn];
        \chainin (nn2) [join= by dspconn];
        \chainin (nn3) [join= by dspconn];
        \chainin (nn4) [join= by dspconn];
        \chainin (nn5) [join= by dspconn];
        \chainin (nn6) [join= by dspconn];
        \chainin (mB) [join= by dspconn];
        \chainin (oC) [join= by dspline];
        \chainin (mult) [join= by dspconn];
        \chainin (shat) [join= by dspconn];

        \chainin (ec0);
        \chainin (mult) [join= by dspconn];

    \end{scope}

\end{tikzpicture}
\vspace{-0.5cm}
\caption{Neural network {\bf architecture of the proposed postfilter} shown in Fig.\,\ref{fig:Figure1}, with Bark-scale mapping (green).}\label{fig:Figure2}
\end{figure}
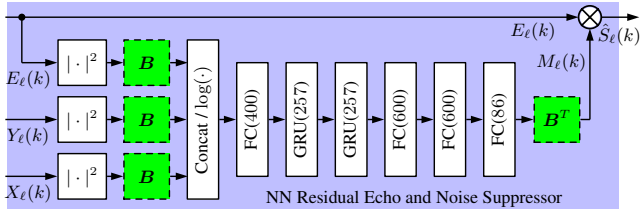

\subsection{Problem Formulation}
\label{ssec:problem}

We consider a speakerphone application used for full-duplex communication as shown in Fig.\,\ref{fig:Figure1}. The device has microphone(s) as well as loudspeaker(s) and is located in a room on a table. In this work, we consider a single microphone and loudspeaker, hence the signal received by the microphone is given by:
\begin{equation}\label{Eq1}
y(n) = s(n) + n(n) + d(n),
\end{equation}
with $s(n)$ as the (desired) near-end speech signal, $n(n)$ denoting the additive background noise at the nearend, $x(n)$ being the farend signal, and $ d(n) = h(n) * f_\mathrm{NL}\left(x(n)\right)$ being the echo. Here, $h(n)$ is the room impulse response (RIR), $f_\mathrm{NL}(\cdot)$ is the nonlinear function modeling the loudspeaker, and $*$ denotes the convolutional operator.\\
\indent The problem formulation is as follows: Given the observed microphone signal $y(n)$ and farend signal $x(n)$, estimate the clean near-end speech signal $s(n)$. With the LEC filter estimate $\hat{h}(n)$, the linear echo component ${d}(n)$ in the microphone signal is approximated by
\begin{equation}\label{Eq2}
\hat{d}(n) = x(n) * \hat{h}(n).
\end{equation}
Subtracting the estimated echo $\hat{D}_\ell(k)={X}_\ell(k)\hat{H}_\ell(k)$ in the discrete Fourier transform (DFT) domain, we obtain
\begin{equation}\label{Eq3}
E_\ell{(k)} = Y_\ell(k) - {X}_\ell(k)\hat{H}_\ell(k),
\end{equation}
where $k$ is the frequency bin index and $\ell$ the frame index. In a hybrid system, the NN RES and noise suppression stages are trained to find the time-frequency mask $M_\ell(k)$ to be applied on the LEC output $E_\ell{(k)}$, yielding the final estimate for the clean nearend speech by
\begin{equation}\label{Eq4}
\hat{S}_\ell(k) = M_\ell(k) E_\ell{(k)}.
\end{equation}
Using \eqref{Eq3}, \eqref{Eq4}, and the DFT of \eqref{Eq1}, we obtain
\begin{equation}\label{Eq5}
\hat{S}_\ell(k)=M_\ell(k)S_\ell(k) + M_\ell(k)\big(N_\ell(k)+\Delta{D}_\ell(k)\big),
\end{equation}
where we defined $\Delta{D}_\ell(k)=H_\ell(k)X_\ell^{\prime}(k)-\hat{H}_\ell(k){X}_\ell(k)$ and further $x^{\prime}(n)=f_\mathrm{NL}\big(x(n)\big)$ and its DFT as $X_\ell^{\prime}(k)$. The overall estimation error term defined as $\epsilon_\ell(k)=\hat{S}_\ell(k)-S_\ell(k)$ is then given by
\begin{equation}\label{Eq6}
\epsilon_\ell(k)\!=\!\big(M_\ell(k)\!-\!1\big)\cdot S_\ell(k)+M_\ell(k)\cdot \big(N_\ell(k)+\Delta{D}_\ell(k)\big).  
\end{equation}
It can be seen that the error consists of two terms: near-end signal distortion and residual noise and echo error, which itself consists of two components: masked noise and the residual echo due to the remaining non-linear echo not addressed by LEC. The task of the neural network is to estimate the clean near-end speech as its target via minimizing the overall estimation error \eqref{Eq6}. 

\subsection{Proposed System}
\label{ssec:LEC}

Fig.\,\ref{fig:Figure1} shows the block diagram for the proposed system, including an LEC block for linear acoustic echo cancellation and an NN postfilter stage. Each stage is described below. The entire model operates (apart from the proposed perceptual mapping) in the DFT domain. To accomplish this, input signals are square-root Hann windowed and the resulting frames are then transformed by a $K$-point DFT. For synthesis of the output signals, we use another square-root Hann window, $K$-point IDFT, and overlap-add.\\
\noindent\textbf{Linear echo canceller}: To assure a minimal aliasing level both in-band and across sub-bands, we use an over-sampled filterbank after \cite{Harteneck1999}. For the echo cancellation adaptive algorithm, we employ a subband NLMS filter with joint optimization on both the normalized step-size and regularization parameters \cite{Benesty2015a}.\\
\noindent\textbf{Proposed neural network postfilter}: The architecture of the proposed neural network used as postfilter in the hybrid system is shown in Fig.\,\ref{fig:Figure2}. We choose an {\tt NSNet2}-like architecture\cite{Braun_NSNET2}, which we consider to be a good balance between achievable performance and computational complexity of the solution. It consists of fully connected (FC) and gated recurrent unit (GRU) layers. Motivated from psycho-acoustics and perceptual relevance of log-Bark power spectral features, we apply a mapping from DFT into Bark domain, denoted by the $K\times B$ matrix $\mathbf{B}=(B(k,b))$, applied on the DFT power spectra of the inputs ($|E_\ell(k)|^2, |Y_\ell(k)|^2, |X_\ell(k)|^2$). The output of the filter with input $X_{\ell}(k)$ is then given by $Z_{\ell}(b)=\sum_{k\in{\mathcal{K}}}{B(k,b)|X_{\ell}(k)|^2}$, with $k\in\mathcal{K}=[0,K\!-\!1]$, where $B(k,b)$ is the $b$th filter computed within the frequency bin range between $f_{\mathrm{l}}(b)$ and $f_{\mathrm{u}}(b)$, which refer to the starting and end band edges in frequency for the $b$th filter. For the $b$th frequency band, the contribution from the energy in the $k$th DFT bin is given by \cite{Kabal_PhDthesis}:
\begin{equation}
B(k,b) \!=\! \frac{\max\!\!\big[0,\!\min\!\big(f_{\mathrm{u}}(b),\!\! \frac{(2k\!+\!1){f_s}}{2K}\!\big)\!-\!\max\!\big(f_{\mathrm{l}}(b),\!\!\frac{(2k\!-\!1){f_s}}{2K}\!\big)\!\big]}{{f_s} / K}
\end{equation}
where $f_s$ is the sampling rate. We design auditory-motivated filters following the Bark scale to uniformly divide the frequency bin range $k\in\mathcal{K}$ into a number of $B$ bands. Finally, these mapped features are concatenated and compressed by a logarithm before passing them to the first FC layer. After returning from the mapped domain using the transpose of the mapping matrix ($\mathbf{B}^T$), the NN yields a real-valued mask $M_\ell(k)$.\\
\noindent\textbf{Loss function:} As training loss, we employ the 
spectral complex compressed mean-squared error (CCMSE) \cite{Ephrat2018} that combines magnitude and phase-aware terms according to
	\begin{equation}
		\label{eq:speechloss}
        \begin{split}
            J^\mathrm{CCMSE} &= \sum_{k,\ell} (1\!-\!\alpha) \big||\tilde{S}_\ell(k)|^c \!-\! |S_\ell(k)|^c \big|^2 \\
            &+ \alpha \big| |\tilde{S}_\ell(k)|^c e^{j\varphi_{\tilde{S}}(\ell,k)} - |S_\ell(k)|^c e^{j\varphi_S(\ell,k)} \big|^2\!\!,
        \end{split}
	\end{equation}
\noindent where $0<\alpha<1$ is a weighting factor between the two terms, \mbox{$c=0.3$}, and $\tilde{S}_\ell(k)$ is obtained from sequence $\hat{s}(n)$, again square-root Hann windowing, and DFT (known as short-term Fourier transform consistency enforcement  \cite{Wisdom2019}).

\section{Training and Evaluation Framework}
\label{sec:Framework}

\subsection{Datasets and Augmentation}
\label{ssec:data}

\noindent \textbf{Training set:} As speech material for $s(n)$ and $x(n)$ we use data provided within the Microsoft Acoustic Echo Cancellation Challenge 2023, consisting of $50,\!000$ recordings from $10,\!000$ environments, recorded by Mechanical Turk users \cite{cutler2023AEC}. We create training signals of 10 s length. As background noise $n(n)$, we include noise files from the training dataset of the ICASSP 2023 Deep Noise Suppression Challenge~\cite{dubey2023}.
Noises are added to the speech at signal-to-noise ratios uniformly distributed with SNR$\sim\!\mathcal{U}[0,30]$ dB. We use a pool of simulated and real RIRs, with various room configurations and isotropic and point-source noises \cite{RIR_SLR}. A random silent period of up to $10$\,s is included for nearend, echo, and noise to add more realism towards the dynamics of a dialogue in a meeting. \\
\indent To account for non-linearities of the loudspeaker \cite{HaoZhang2022}, $80\%$ of the files of $x^{\prime}(n)$ include $f_\mathrm{NL}(x(n))$ either as error function \mbox{$x^{\prime}(n)={\eta^{-1}}{\erfc\big({\eta}\cdot x(n)\big)}$} with $\eta=1$, or as a scaled version $\big(0\!<\!\eta\!<\!1 \text{ if } x(n)\!\!<\!\!0,\text{ else }\eta\!=\!1\big)$ with $\eta\!\sim\!\mathcal{U}[-12,0]$ dB, applied to the negative parts of the reference signal $x(n)$ to reflect the non-linearities of the transducer in a product. 
The echo $d(n)$ is added with a signal-to-echo ratio SER$\,\sim\!\mathcal{U}[-30,10]$ dB. As the original data is all fullband, we choose $32$-bit linear PCM when writing the audio files. We convert the sampling frequency of all data to 16 kHz. {To simulate clock drifts due to resampling, we consider a 1\% drift of the device's nominal sampling rate. We augment the echo signal generation with cross-fading between two RIRs. To simulate close distance between speaker and microphone as well as volume change, the second RIR is adjusted in its direct path energy via a gain following a standard deviation of $\sigma=1.0$ as dynamic RIR mixing gain, and then added to the first RIR. Random bandpass filtering is applied on signal components to reflect the device's frequency response and any mismatch in the device.}\\
\noindent \textbf{Test set}: The ICASSP 2023 AEC Challenge blind test set \cite{cutler2023AEC} is used to evaluate the performance of the methods studied in this work. It consists of $800$ real-world clips with $300$ double-talk (DT), $300$ single-talk farend (STFE), and $200$ single-talk nearend (STNE) examples. The audio files are of variable length, ranging between $30$ to $45$ seconds. They cover a large variety of distortion types, including strong speaker/mic distortions, stationary/non-stationary noise, glitches resulting in chippy audio, varying gain during the recording, and a cascade effect due to audio processing of the device being active. Please note that the STFE subset of the blind test set consists of very difficult delay estimation cases---either long or variable delays aside from non-linear distortions and stationary noise scenarios \cite{cutler2023AEC}. However, as the delay between the mic and loudspeaker mounted on the same device is a known parameter, such 
long-delay and variable jitter conditions are no typical use case for a speakerphone's echo control processing, hence they are not the main focus for our target application in this study. Accordingly, to exclude the impact of the existing jitter/delay between the farend signal and mic signal in the test set, we performed offline delay compensation of the test data, using cross-correlation between $y(n)$ and $x(n)$. 

\subsection{Training Setup} 
\label{ssec:setup}

As LEC we use the normalized least mean-squares (NLMS) method implemented with an over-sampled filterbank \cite{Shachar2023}. A sampling rate of $16$\,kHz, window length of $1024$ samples, DFT size of $K=512$ samples, and a frame shift of 128 samples are used with the prototype filter designed as proposed in \cite{Harteneck1999}. Four filter lengths of $N_\mathrm{LEC}\in\{4,8,16,32\}$ taps are considered to address both fast and slow reactions to echo path changes common in realistic conditions as well as to adjust the adaptation speed as a function of the adaptive filter length accordingly. To achieve a better generalization and performance of the LEC stage on mismatched test sets, we allowed LEC parameters to vary randomly in training \cite{panchapagesan23_interspeech}; namely both, adaptive filter lengths and the smoothing coefficient $\beta\!\sim\!\mathcal{U}[0.5,2]$ for PSD estimation, as factor variation. Motivated from perceptual audio coding and perceptual audio measurement, the postfilter Bark scale is chosen to decompose DFT bins into $B=86$ uniformly spaced bands on the Bark scale covering a frequency range of 0 Hz to 8 kHz (for design details, we refer the reader to \cite{Kabal_PhDthesis}).\\
\noindent \textbf{Network Training:} We initialize the learning rate to $10^{-4}$, which drops by a factor of $0.5$ once the validation metric does not improve for $10$ epochs. We trained the models with $400$ epochs, whereby one training epoch comprises $95,\!000$ sequences. For gradient calculation, the Adam optimizer \cite{Adam_reference} is used in its standard configuration. All models are implemented and trained using PyTorch.

\subsection{Reference Methods} 
\label{ssec:reference}

As reference methods, we consider the following methods:  (i) LEC stage-only, (ii) the fully data-driven {\tt DTLN} model\footnote{Retrained on our own training data.} \cite{westhausen21_icassp} with 4 consecutive LSTM layers with 256 units each, followed by a fully connected layer with sigmoid activation function, and (iii) the recently proposed efficient {\tt DeepVQE-S} \cite{indenbom2023deepvqe} (denoted as {\tt DVQE-S}) as the state of the art used in Microsoft Teams\footnote{Here, our own implementation is retrained on our training data. Note that in our reference {\tt DVQE-S} model, we drop the self-attention layer for signal alignment, but rather focus on evaluation of the architecture used for echo control, allowing to demonstrate the achievable performance of the neural network architectures and not its ability to solve the delay compensation or synchronization as detailed in \cite{indenbom2023deepvqe,sca,indenbom2023deep}.}, and finally, (iv) we include the results obtained by our proposed method without perceptual mapping \textbf{B}, i.e., DFT log-power features as inputs following exactly the same NN topology shown in Fig.~\ref{fig:Figure2}, loss function, and consistency. This will highlight the impact of the proposed perceptual mapping used in the proposed model.

\subsection{Evaluation Metrics} 
\label{ssec:metrics}

For the instrumental evaluation, we mainly use the AECMOS metric, a non-intrusive, model-based echo cancellation quality predictor~\cite{aecmos_2022}. AECMOS individually evaluates echo suppression effectiveness and the effects of other degradations (e.g., noise, nearend degradation). Both measures are reported for DT condition (labeled DT Echo and DT Other).  In the STFE condition, only ST Echo is considered. Here, we also report the average logarithmic echo return loss enhancement (ERLE) \cite{EnznerBuchner_AcousticEchoControl}, given by \mbox{ERLE $= 10 \log_{10}\big(\mathbin{\|{y(n)}\|_2^2/\|{\hat{s}(n)}\|_2^2}\big)$},
which measures the echo reduction between the unprocessed and enhanced signals when only noise and farend signal are present. Finally, to evaluate the denoising capacity of the methods in the STNE condition, we use ST Other and the DNSMOS metrics, following ITU-T P.835, focused on speech distortion (SIG), background noise attenuation (BAK), and overall quality (OVRL) \cite{DNSMOS}.

\section{Results and Discussion}
\label{sec:Results}

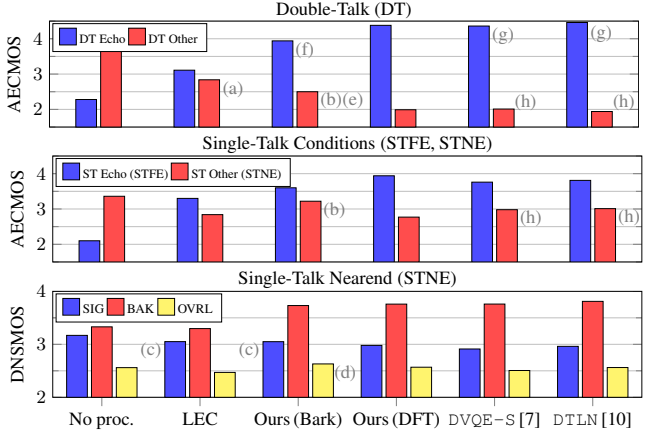
\begin{figure}[t]
 \hspace{-0.125cm}
	\small
    \newlength\figureheight
    \newlength\figurewidth
    \setlength\figureheight{1.8cm}
    \setlength\figurewidth{10cm}

    \definecolor{mycolor1}{rgb}{0.749019622802734,0,0.749019622802734}%
    \begin{tikzpicture}
        \begin{axis} [
        ybar,
        scale only axis,
        width=\figurewidth,
        height=\figureheight,
        xmin=0.5,
        xmax=6.5,
        ymin=1.5,
        ymax=4.5,
        ymajorgrids,
        yminorgrids,
        minor y tick num=1,
        xtick=data,area legend, 
        xmajorticks=false,
        title=Double-Talk (DT),
        title style={at={(0.5,0.85)}},
        ylabel={AECMOS},
        legend style={nodes={scale=0.68}, at={(0.005,0.98)}, anchor=north west, legend columns=2}]
            \addplot [fill=blue!70]
            coordinates {
                (1,2.28) 
                (2,3.11) 
                (3,3.94) 
                (4,4.38) 
                (5,4.36) 
                (6,4.46) 
            };
            \addlegendentry{DT Echo};
    
            \addplot [fill=red!70]
            coordinates {             
                (1,3.879) 
                (2,2.838) 
                (3,2.50)  
                (4,1.99)  
                (5,2.01)  
                (6,1.94)  
            };
            \addlegendentry{DT Other};

            \draw (axis cs:2.37, 2.58) node (t) {\color{black!50}(a)};
            \draw (axis cs:3.37, 2.27) node (t) {\color{black!50}(b)};
            \draw (axis cs:3.59, 2.27) node (t) {\color{black!50}(e)};
            \draw (axis cs:3.1, 3.68) node (t) {\color{black!50}(f)};
            \draw (axis cs:5.12, 4.1) node (t) {\color{black!50}(g)};
            \draw (axis cs:5.35, 2.2) node (t) {\color{black!50}(h)};
            \draw (axis cs:6.12, 4.15) node (t) {\color{black!50}(g)};
            \draw (axis cs:6.35, 2.2) node (t) {\color{black!50}(h)};
            
        \end{axis}
    
    \end{tikzpicture}%
 \\
	\small
    \newlength\figureheightB
    \newlength\figurewidthB
    \setlength\figureheightB{1.8cm}
    \setlength\figurewidthB{10cm}

    \definecolor{mycolor1}{rgb}{0.749019622802734,0,0.749019622802734}%
    \begin{tikzpicture}
        \begin{axis} [
        ybar,
        scale only axis,
        width=\figurewidthB,
        height=\figureheightB,
        xmin=0.5,
        xmax=6.5,
        ymin=1.5,
        ymax=4.5,
        ymajorgrids,
        yminorgrids,
        minor y tick num=1,
        xtick=data,area legend, 
        xmajorticks=false,
        title={Single-Talk Conditions (STFE, STNE)},
        title style={at={(0.5,0.85)}},
        ylabel={AECMOS},
        legend style={nodes={scale=0.68}, at={(0.005,0.98)}, anchor=north west, legend columns=2}]
        
            \addplot [fill=blue!70]
            coordinates {
                (1,2.1) 
                (2,3.3) 
                (3,3.6) 
                (4,3.94) 
                (5,3.76) 
                (6,3.81) 
            };
            \addlegendentry{ST Echo (STFE)};
    
            \addplot [fill=red!70]
            coordinates {             
                (1,3.36) 
                (2,2.84) 
                (3,3.22) 
                (4,2.77) 
                (5,2.98) 
                (6,3.01) 
            };
            \addlegendentry{ST Other (STNE)};

            \draw (axis cs:3.37, 2.97) node (t) {\color{black!50}(b)};
            \draw (axis cs:5.37, 2.725) node (t) {\color{black!50}(h)};
            \draw (axis cs:6.37, 2.75) node (t) {\color{black!50}(h)};
            
        \end{axis}
    
    \end{tikzpicture}%
 \\
	\small
    \newlength\figureheightC
    \newlength\figurewidthC
    \setlength\figurewidthC{1.8cm}
    \setlength\figureheightC{10cm}

    \definecolor{mycolor1}{rgb}{0.749019622802734,0,0.749019622802734}%
    \begin{tikzpicture}
        \begin{axis} [
        ybar,
        scale only axis,
        width=\figureheightC,
        height=\figurewidthC,
        xtick={1,2,3,4,5,6},
        xticklabels={No proc., LEC, Ours\,(Bark), Ours\,(DFT), {\tt DVQE-S}\,[7], {\tt DTLN}\,[10]},
        ytick={2,3,4},
        xmin=0.5,
        xmax=6.5,
        ymin=2,
        ymax=4.0,
        ymajorgrids,
        yminorgrids,
        minor y tick num=1,
        xtick=data,area legend, 
        xtick pos=left,
        title=Single-Talk Nearend (STNE),
        title style={at={(0.5,0.85)}},
        ylabel={DNSMOS},
        legend style={nodes={scale=0.7}, at={(0.005,0.98)}, anchor=north west, legend columns=3}]
        
            \addplot [fill=blue!70]
            coordinates {
                (1,3.169) 
                (2,3.051) 
                (3,3.05) 
                (4,2.98) 
                (5,2.912) 
                (6,2.962) 
            };
            \addlegendentry{SIG};
    
            \addplot [fill=red!70]
            coordinates {             
                (1,3.33) 
                (2,3.297) 
                (3,3.732) 
                (4,3.759) 
                (5,3.76) 
                (6,3.811) 
            };
            \addlegendentry{BAK};

            \addplot [fill=yellow!70]
            coordinates {             
                (1,2.56) 
                (2,2.47) 
                (3,2.63) 
                (4,2.568) 
                (5,2.506) 
                (6,2.562) 
            };
            \addlegendentry{OVRL};

            \draw (axis cs:2.5, 2.88) node (t) {\color{black!50}(c)};
            \draw (axis cs:1.5, 2.88) node (t) {\color{black!50}(c)};
            \draw (axis cs:3.48, 2.45) node (t) {\color{black!50}(d)};
            
        \end{axis}
    
    \end{tikzpicture}%
 

 \vspace{-0.3cm}
  \caption{(Top) AECMOS measurements for the \textbf{DT} condition, (middle) ST Echo for the \textbf{STFE} and ST Other for the \textbf{STNE} condition, (bottom) DNSMOS (SIG/BAK/OVRL) reported for the \textbf{STNE} condition. The labels (a) to (h) refer to the discussion below.}\label{fig:Allresults}
  \vspace{-0.4cm}
\end{figure}

\noindent \textbf{AECMOS and DNSMOS results}: We report the averaged instrumental objective measures of AECMOS and DNSMOS obtained for the three conditions (DT, STFE, STNE) of our test set described in Section~\ref{ssec:data}. The results are shown in Fig.~\ref{fig:Allresults}.
We can see that the proposed Bark-scaled hybrid solution yields an overall improvement over the LEC performance in all conditions predicted by AECMOS, except from near-end quality (\mbox{DT Other}), where (a) LEC expectedly scores the highest of all models. Moreover, the comparison against our reference model without Bark transformation shows (b) the immense benefits of the perceptually motivated mapping for nearend speech preservation (DT/ST Other). 
In STNE, our proposed model shows (c) no degradation of SIG when compared to the LEC stage and (d) the highest OVRL score. For the double-talk condition, all the methods offer a trade-off between near-end speech quality (DT Other) and echo removal (DT Echo). 
Apart from the LEC, our Bark-scale method offers (e) the highest \mbox{DT Other} at the expense of (f) a reduced, but still high \mbox{DT Echo}. The (g) good echo removal offered by our reference models comes at the cost of (h) a noticeably worse nearend speech quality in both double-talk (DT Other) and STNE conditions (\mbox{ST Other}). {\it All in all, we can conclude that our hybrid proposal keeps state of-the-art model performance and, through the use of an auditory filterbank, is even capable of achieving a more favorable trade-off regarding nearend speech quality.}\\
 
\noindent \textbf{Echo Return Loss Enhancement}: We further report the echo control and noise reduction performance in STFE condition measured by the ERLE scores, shown in Table~\ref{ERLETable}.
We can see that our proposed NN postfilter significantly improves the echo suppression level achieved by its LEC stage. The {\tt DTLN} model achieves the highest ERLE, followed by our DFT and Bark models, which score comparably despite their strong performance differences reported in Fig.~\ref{fig:Allresults}. {\tt DVQE-S} scores significantly lower. Comparing Fig.\,\ref{fig:Allresults} and Table\,\ref{ERLETable}, we observe that ERLE and ST Echo do not deliver the same rank orders among the methods, e.g., our proposed Bark-scale model scores significantly higher in ERLE than {\tt DVQE-S} despite having a comparable ST Echo score. We associate this observation to the fact that ST Echo reflects the subjective impression about echo removal \cite{cutler2021crowdsourcing}, while ERLE simply measures the residual signal's energy. AECMOS might overlook residual echo if it is no longer speech-like, as the underlying NN was trained on subjective listening tests, where noise-like residual echo would be rated as less annoying or simply be not associated with the echo component anymore. ERLE, on the other hand, punishes each type of residual echo and also considers noise suppression effectiveness.\\ 

\begin{table}[t]
\caption{\textbf{ERLE in dB} for the \textbf{STFE} condition, averaged over the respective 300 files of the test set.}\label{ERLETable}
\vspace{-0.25cm}
\begin{center}
\scalebox{0.95}{
\begin{tabular}{cccccc} 
\toprule
{LEC} & Ours (Bark)  & Ours (DFT)  & {\tt DVQE-S}\!\cite{indenbom2023deepvqe} & {\tt DTLN}\!\cite{westhausen21_icassp}  \\ 
\midrule 
 37.57 & {60.10} & \underline{62.00} & 40.00 & \textbf{68.78} \\
 \bottomrule
\end{tabular}}
\end{center}
\vspace{-0.7cm}
\end{table} 

\begin{table}[t]
\caption{\textbf{Efficiency analysis} for the studied methods. Lowest demand is marked \textbf{bold} and second best results are \underline{underlined}.}\label{FootprintTable}
\vspace{-0.25cm}
\begin{center}
\scalebox{0.86}{
\begin{tabular}{l rrrrr}  
\toprule
Attribute  & Ours (Bark) & Ours (STFT) & {\tt DVQE-S}\!\cite{indenbom2023deepvqe} & {\tt DTLN}\!\cite{westhausen21_icassp}\\
\midrule
Param (M) & \underline{1.58} & 2.04 & \textbf{0.72}  & {3.16}\\ 
MACs/s (M)  & \textbf{235.00}  & \underline{240.00} & 2170.00 & 408.00 \\ 
RTF (\%) & \underline{0.22} & {0.23} & \textbf{0.20} & 0.97 \\
\bottomrule
\end{tabular}}
\end{center}
\vspace{-0.6cm}
\end{table}


\noindent \textbf{Efficiency analysis}: Table~\ref{FootprintTable} shows the memory footprint and complexity analysis\footnote{We used \url{https://pypi.org/project/torchinfo/}.} of the methods studied in this work, entailing number of multiply-accumulate operations per second (MACs/s), the number of trainable parameters, and the realtime factor (RTF), measured on an {\tt Intel i9-10850K} CPU at 3.60 GHz.
We observe that {\tt DVQE-S} has the lowest parameter count and RTF, followed by our proposed Bark model. Both architectures prove realtime-capable on the tested CPU. However, since {\tt DVQE-S} mainly consists of convolutional layers, way more MACs/s are required. In contrast, {\it our proposed fully connected model only requires about 10\% of the {DVQE-S} models' MACs/s}. Moreover, on a device like a speakerphone, efficiently implementing convolutional architectures is much more difficult than it is for fully connected ones. This means that RTF rankings can change significantly depending on the deployment platform. While our model has a higher parameter count than {\tt DVQE-S}, the difference is not critical on modern devices. Combined with the previously reported good performance, {\it the low complexity of our model makes it a great fit for application in speakerphones}.

\section{Conclusions}
\label{sec:Conclusion}
\vspace{-0.2cm}
In this paper, we presented a realtime-capable joint acoustic echo control and noise reduction model for speakerphones, consisting of a linear echo canceller, followed by a lightweight neural network as postfilter. Our results demonstrated that the proposed design achieves an on-par or even improved performance compared to existing state-of-the-art solutions. An attractive trade-off between echo suppression and near-end speech preservation is offered along with a reasonable amount of model parameters. {\it Our proposed model excels all reference methods by far in computational complexity}.
\ninept
\bibliographystyle{IEEEbib}
\bibliography{IEEEabrv,MGCabrv,strings,refs}

\end{document}